
\magnification=\magstep1
\baselineskip=20pt
\parskip2mm
\input amssym.def
\input amssym

\def\sl{sl(2,{\Bbb C})}
\def\slq{sl_q(2,{\Bbb C})}
\def\Sl{Sl(2,{\Bbb C})}
\def\bP{{\Bbb P}^1}
\def\bPP{\bP \coprod {\bP}^*}

SISSA-58/93/FM

\vskip3truecm
\centerline{\bf A Borel-Weil-Bott approach to}

\centerline{representations of $\slq$}

\vskip3truecm
\centerline{Davide Franco and Cesare Reina}

\centerline{SISSA - Strada Costiera 11 - TRIESTE (Italy)}

\vfill{\centerline{FRANCOD@ITSSISSA - REINA@ITSSISSA}}
\eject

In this letter we use a simple realization of the
quantum group $\slq$, which does not seem to be widely known.
As we shall see, the advantage is that we can give a geometric
interpretation of the representations of $\slq$ much on the lines
of the standard Borel-Weil-Bott theory
(see e.g. [W] for an introduction).
A  purely algebraic approach can be found in [BL].

As well known, the classical group $\Sl$ naturally acts on the
projective space ${\Bbb P}(V)$ of the fundamental representation.
This action is induced by the natural action on $V\simeq {\Bbb C}^2$.
By pull-back, one gets an action on the space of holomorphic
functions $f:{\Bbb C}^2\rightarrow {\Bbb C}$. Let
$z_1,z_2\in {\Bbb C}^2-\{ 0\}$ be
the homogeneous coordinates on $\bP$. The infinitesimal
action of the standard generators induces the fundamental vector fields
$$H=z_1\partial_{z_1}-z_2\partial_{z_2},\qquad
  E=z_1\partial_{z_2},\qquad F=z_2\partial_{z_1},$$
which obviously realize the algebra of $\sl$.
The eigenvectors of $H$ are obviously homogeneous polynomials
$f=f(z_1,z_2)$. In particular, $z_1^n$ is the highest weight vector
of weight $n$ and the spaces $S_n$
of homogeneous polynomials of degree $n$ realize the
irreducible  representation
of spin $n/2$.
The geometrical interpretation of the irreducible representations
realized in this form
is the simplest example in BWB theory; one finds that $S_n$ is the space
$H^0(\bP ,H^n)$ of the holomorphic sections of the $n$-th power
of the hyperplane bundle $H$ (see e.g. [FH]).

In the following we extend such a description to the quantum
group $\slq$. We start with
$$H_1=\left( \matrix{1&0\cr 0&0\cr}\right) ,\qquad
  H_2=\left( \matrix{0&0\cr 0&1\cr}\right) ,$$
as generators of the centre of $gl(2,{\Bbb C})$, the Cartan
generator of $\sl$ being $H=H_1-H_2$. We exponentiate these
generators to $e^{tH_i},\; (t\in {\Bbb C})$ and set
$e^t=q\in {\Bbb C}^*$. As $q$ varies in ${\Bbb C}^*$, $q^H$ varies
in the full Cartan subgroup.
The natural action on ${\Bbb C}^2$
induces the pull-back action on functions by
$$\eqalign{&(q^{H_1}f)(z_1,z_2)=f(z_1q,z_2)\cr
           &(q^{H_2}f)(z_1,z_2)=f(z_1,z_2q)\cr
           &(q^{H}~f)(z_1,z_2)=f(z_1q,z_2q^{-1})\cr}$$
One knows that, setting
$$\hat K=q^H,\qquad \hat E=z_1D_2,\qquad \hat F=z_2D_1,$$
where
$$D_i={{q^{H_i}-q^{-H_i}}\over {z_i(q-q^{-1})}}$$
is the finite difference operator in the $i$-th coordinate,
one gets that the quantum group relations of $\slq$ are satisfied.
Indeed, one easily checks that $\hat E$ ($\hat F$) is an
eigenvector for the adjoint action of $\hat K$ with eigenvalue
$q^2$ $(q^{-2})$ (notice that the exponents are actually
the roots of $\sl$). A short computation yields the other
relation
$$[z_1D_2,z_2D_1]={{\hat K-\hat K^{-1}}\over {q-q^{-1}}}.$$

Let us now look for a realization of the finite dimensional
irreducible representations. Since $\hat K=q^H$,
any eigenvector of $H$ with weight $m$ is an eigenvector for
$\hat K$ as well with weight $q^m$. As $\hat E z_1^n=0$,
it follows that the spaces of the irreducible representations
coincide with the classical ones $S_n$; a well known fact.
Notice that, if one thinks of $q$ as a fixed parametes,
one has to stay far from roots of unity
to get that $\hat K$ gives a sensible weight space decomposition
of any irreducible representation. However this is not a true
problem for $\slq$, because if one keeps $q$ variable one can go
to roots of unity by continuity, getting weight space decompositions
even there.

The space $S=\oplus _{n>0} S_n$ is a graded ring, to
which one canonically associates a projective scheme (see e.e. [H]).
In our case, this is the projective line
$\bP \simeq Proj(S)$ and we get an isomorphism $S_n=H^0(\bP ,H^n)$.
Although the finite difference operators $D_i$ can be understood
as acting on $\bP$ by pull-backs, they are not differential
operators. To get a "differential version" of their action, we
introduce a (non-)commutative version of $\bP$, following
the ideas of [C] (see also [VGB] for a short introduction to
"double spaces").

We take the abelian algebra ${\cal A}=S\oplus S^*$, where $S^*$ is
the symmetric tensor algebra over $V$. There is again a scheme associated
to $\cal A$, which is actually the disjoint union $\bPP$.
On this scheme we have a natural action of $\hat K$ given by
$$\hat K(z_1,z_2)=q^H(z_1,z_2)=(z_1q,z_2q^{-1})$$
on the first $\bP$ and by
$$\hat K(w_1,w_2)=q^{-H}(w_1,w_2)=(w_1q^{-1},w_2q)$$
on the second one. This action extends to $\cal A$ as
$$\hat K_{\cal A}(f_1,f_2)=(q^Hf_1, q^{-H}f_2).$$

The standard differential operator
$d:{\cal A}\rightarrow \Omega ^1({\cal A})$ reads
$$d(f_1,f_2)=(f_2-f_1,f_1-f_2)$$
and the elements of the representation spaces $S_n$ live in the
kernel of $d$, once one identified $\bP$ with ${\bP}^*$ thanks to
the isomorphism $V\simeq V^*$. Now, the difference operators read
$$D_i={1\over {z_i(q-q^{-1})}}d\circ \hat K_{\cal A}$$
and one easily checks that they are derivations of $\cal A$.

Summing up, we see that $\bPP =Proj({\cal A})$ works as a
"homogeneous space" for $\slq$ as in standard BWB theory.

A similar construction can be done for $sl_q(3,{\Bbb C})$, as
far as the irreducible representations with highest weights on
the walls of the Weyl chamber are concerned. The generic
representations deserve more work.

\vskip7truecm
\centerline {\bf References}
\vskip3truemm
\item{[BL]}{ Biedenharn L.C., Lohe M.A.: Comm. Math. Phys. 146 (92) 483}
\item{[C]}{ Connes A.: "Geometrie non commutative" InterEditions, Paris, 1990 }
\item{[FH]}{ Fulton W. and Harris j. "Representation theory" GTM -
Springer (1991)}
\item{[H]}{ Hartshorne R.: "Algebraic Geometry" Springer G.T.M. }
\item{[VGB]}{ Varilly J.C., Gracia-Bondia J.M.: " Connes' Noncommutative
Differential Geometry and the Standard Model" preprint }
\item{[W]}{ Wallach N.R.: "Harmonic analysis on homogeneous spaces" Dekker, New
Yor, 1973 }
\bye